\documentclass{ws-ijmpe}
\usepackage{amssymb}

\usepackage{amsmath}

\begin{document}

\markboth{Hilmar Forkel, Michael Beyer and Tobias Frederico}{Linear Meson
and Baryon Trajectories in AdS/QCD}

\catchline{}{}{}{}{} 

\title{LINEAR MESON AND BARYON TRAJECTORIES IN ADS/QCD }
\author{{\footnotesize HILMAR FORKEL} }
\address{Departamento de F\'isica, ITA-CTA, 12.228-900 S\~ao Jos\'e dos Campos, S\~ao
Paulo, Brazil and Institut f\"ur Theoretische Physik, Universit\"at
Heidelberg, D-69120 Heidelberg, Germany}

\author{{\footnotesize MICHAEL\ BEYER }}
\address{Institut f\"ur Physik, Universit\"at Rostock, D-18051 Rostock,
Germany}

\author{{\footnotesize TOBIAS FREDERICO}}
\address{Departamento de F\'isica, ITA-CTA, 12.228-900 S\~ao Jos\'e dos Campos, 
S\~ao Paulo, Brazil}

\maketitle

\begin{history}
\received{(received date)}
\revised{(revised date)}
\end{history}\bigskip

\begin{abstract}
An approximate holographic dual of QCD is constructed and shown to reproduce
the empirical linear trajectories of universal slope on which the square
masses of radially and orbitally excited hadrons join. Conformal symmetry
breaking and other IR effects are described exclusively by deformations of
the anti-de Sitter background metric. The predictions for the light hadron
spectrum include new relations between ground state masses and trajectory
slopes and are in good overall agreement with experimental data.
\end{abstract}\bigskip

The AdS/QCD program aims at constructing the holographic dual of Quantum
Chromodynamics (QCD) on the basis of the AdS/CFT correspondence\cite{mal97}
and experimental data. Since the pioneering work\cite{pol02}, this approach
has met with considerable success in describing vacuum and hadron properties
by a local field theory for the QCD dual in IR-deformed AdS$_{5}$ spacetime
(and potentially other background fields) whose form and parameters are
constrained by experimental information. For various investigations in this
direction see e.g. the references in\cite{for07}.

Much of this work was based on the minimal ``hard wall'' implementation of
IR effects and confinement\cite{pol02}. Although very useful in several
respects, this abrupt compactification of the fifth dimension predicts
quadratic instead of linear square mass trajectories both as a function of
spin and radial excitation quantum numbers\cite{det05,kat06}, however, in
contrast to experimental data and semiclassical string model arguments\cite%
{shi05}.

Different ways to overcome this problem in the meson sector have been
proposed in Refs.\cite{kar06,and06}, and a dual description for the
similarly pronounced empirical trajectories\cite{kle02} in the baryon sector
has been found recently\cite{for07}. The latter reproduces the linear
trajectories\cite{kle02,ani00} $M^{2}=M_{0}^{2}+W\left( N+L\right) $ of
Regge type for radial ($N$) and orbital ($L$) excitations in both meson and
baryon spectra. The holographic description relies on the identification of high
angular momentum states with metric fluctuations\cite{det05,bro04} and
naturally accommodates the approximately universal empirical slope $W\sim 1.1
$ GeV$^{2}$ of all trajectories\cite{for07}.

For an UV conformal gauge theory like QCD, the dual string spacetime is the
product of a five-dimensional non-compact part which asymptotically becomes
anti--de Sitter space $\mathrm{AdS}_{5}\left( R\right) $ of curvature radius 
$R$, and a five-dimensional compact Einstein space $X_{5}$. The line element
therefore takes the form\cite{pol02}%
\begin{equation}
ds^{2}=e^{2A\left( z\right) }\frac{R^{2}}{z^{2}}\left( \eta _{\mu \nu
}dx^{\mu }dx^{\nu }-dz^{2}\right) +R^{2}ds_{X_{5}}^{2}  \label{metric}
\end{equation}%
where $\eta _{\mu \nu }$ is the four-dimensional Minkowski metric and $%
A(z)\rightarrow 0$ as $z\rightarrow 0$ in order to reproduce the conformal
behavior of asymptotic freedom at high energies. The string modes $\phi
_{i}\left( x,z\right) =e^{-iP_{i}x}f_{i}\left( z\right) $ dual to physical
states of the gauge theory are particular solutions of the wave equations in
the geometry (\ref{metric}) and fluctuations around it\cite{mal97}.
Rewritten as eigenvalue problems, the field equations become 
\begin{equation}
\left[ -\partial _{z}^{2}+V_{M}\left( z\right) \right] \varphi \left(
z\right) =M_{M}^{2}\varphi \left( z\right)  \label{eveqm}
\end{equation}%
for the normalizable string modes $\varphi \left( z\right) =g\left( z\right)
f_{M}\left( z\right) $ dual to spin-0, 1 mesons as well as 
\begin{equation}
\left[ -\partial _{z}^{2}+V_{B,\pm }\left( z\right) \right] \psi _{\pm
}\left( z\right) =M_{B}^{2}\psi _{\pm }\left( z\right)  \label{eveqb}
\end{equation}%
for the string modes $\psi _{\pm }\left( z\right) =h\left( z\right) f_{B,\pm
}\left( z\right) $ dual to spin-1/2, 3/2 baryons ($\pm $ denote the fermion
chiralities)\cite{det05}. The eigenvalues $M_{M,B}^{2}$ give the hadronic
mass spectrum and the boundary conditions on the eigensolutions $\varphi $
and $\psi _{\pm }$ determine the corresponding gauge theory operator\cite%
{mal97}. The duals of hadron states $\left| i\right\rangle $ behave as $%
f_{i}\left( z\right) \overset{z\rightarrow 0}{\longrightarrow }z^{\tau _{i}}$
where $\tau _{i}=\Delta _{i}-\sigma _{i}$ is the twist of the
lowest-dimensional gauge theory operator which creates $\left|
i\right\rangle $. (Thereby $\tau _{i}$ enters the mass terms of the
potentials $V_{M,B}$.) The lightest string modes are then associated with
the leading twist operators, and therefore with the valence quark content of
the low-spin (i.e. spin 0, 1/2, 1, and 3/2) hadron states\cite{det05}. The
duals of orbital excitations are identified with fluctuations about the AdS
background\cite{det05,bro04}.

The first part of our strategy is to find a minimal modification of the
potentials $V_{M,B}$ in the AdS$_{5}$ background (given by Eq. (\ref{metric}%
) with $A\left( z\right) \equiv 0$) which reproduce the desired linear
trajectories with universal slopes. It turns out that this requires to break
conformal symmetry via the replacement $\tau _{i}\rightarrow \tau
_{i}+\lambda ^{2}z^{2}$ in the AdS$_{5}$ potentials. With $\tau _{M}=L+2$
for the twist of the meson interpolators the mesonic potential becomes%
\begin{equation}
V_{M}^{\left( \text{LT}\right) }\left( z\right) =\left[ \left( \lambda
^{2}z^{2}+L\right) ^{2}-\frac{1}{4}\right] \frac{1}{z^{2}}  \label{vMconf}
\end{equation}%
while the baryon potential, associated with the\ interpolators of twist $%
\tau _{B}=L+3$, turns into 
\begin{equation}
V_{B,\pm }^{\left( \text{LT}\right) }\left( z\right) =\left\{ \left(
L+1\right) \left( L+1\mp 1\right) +\left[ 2\left( L+1\right) \pm 1\right]
\lambda ^{2}z^{2}+\lambda ^{4}z^{4}\right\} \frac{1}{z^{2}}.  \label{vBconf}
\end{equation}%
The eigenfunctions\cite{for07} have appreciable support only in the small
region $z\lesssim \sqrt{2}\lambda ^{-1}\simeq \Lambda _{\text{QCD}}^{-1}$
close to the UV brane, which is an expected consequence of confinement. The
corresponding eigenvalues 
\begin{equation}
M_{M}^{2}=4\lambda ^{2}\left( N+L+\frac{1}{2}\right) ,\text{ \ \ \ \ \ }%
M_{B}^{2}=4\lambda ^{2}\left( N+L+\frac{3}{2}\right) ,  \label{bspec}
\end{equation}%
show that the square masses of both mesons and baryons are indeed organized
into the observed $N+L$ trajectories. Moreover, Eqs. (\ref{bspec}) predict
the universal slope 
\begin{equation}
W=4\lambda ^{2}  \label{slope}
\end{equation}%
for both meson and baryon trajectories in terms of the IR scale $\lambda $
and exhibit a mass gap of order $\sqrt{W}$. The intercepts $M_{i,0}^{2}$
relate the trajectory slope in a new way to their ground state masses,%
\begin{equation}
M_{M,0}^{2}=\frac{W}{2},\text{ \ \ \ \ \ }M_{B,0}^{2}=\frac{3W}{2}.
\label{mwbrel}
\end{equation}%
In order to show how the above potentials (\ref{vMconf}), (\ref{vBconf})
arise from a holographic dual by specific IR deformations of AdS$_{5}$, we
now construct the corresponding background metric explicitly. This can be
done by equating the field mode potentials in the geometry (\ref{metric})
with \textit{a priori} undetermined warp functions $A\left( z\right) $ to
their counterparts (\ref{vMconf}), (\ref{vBconf}) and leads to the
nonlinear, inhomogeneous differential equation%
\begin{equation}
\pm \left( zA^{\prime }-1\right) +le^{A}-\left[ l\left( l\mp 1\right)
+\left( 2l\pm 1\right) \lambda ^{2}z^{2}+\lambda ^{4}z^{4}\right] \left(
le^{A}\right) ^{-1}=0  \label{bode}
\end{equation}%
of first order (where $l\equiv L+1$ and $\pm $ denotes the chirality) in the
baryon sector and to the second-order equation 
\begin{equation}
z^{2}A^{\prime \prime }+\frac{3}{2}\left( zA^{\prime }\right)
^{2}-3zA^{\prime }+\frac{2}{3}\left( L^{2}-4\right) \left( e^{2A}-1\right) -%
\frac{2}{3}\lambda ^{2}z^{2}\left( \lambda ^{2}z^{2}+2L\right) =0
\label{mde}
\end{equation}%
for the mesons. Of course, it is \textit{a priori} uncertain whether there
exists an approximate gravity dual whose IR deformation can reproduce a
given five-dimensional potential and spectrum, simply because it may not
result from a boundary gauge theory. This is reflected in the fact that the
differential equations for $A\left( z\right) $ may not have physically
acceptable solutions subject to the conformal boundary condition $A\left(
0\right) =0$.

In our case, however, such solutions exist. The exact solution of Eq. (\ref%
{bode}) for both baryon chiralities can even be found analytically,%
\begin{equation}
A_{B}\left( z\right) =\ln \left( 1+\frac{\lambda ^{2}z^{2}}{L+1}\right) .
\label{bsoln}
\end{equation}%
The solutions $A_{M}$ of Eq. (\ref{mde}) are more multi-faceted and show new
properties, including dynamical compactification of the $z$ direction (a
dual signature for confinement) for the low-lying orbital excitations due to
tachyonic (but stable) string modes. Those and other features, including the
preservation of linear trajectories\ even in the compactified $z$ interval
and the origin of the $L$ dependence, are discussed and compared to other
holographic duals for linear meson trajectories in Ref.\cite{for07}.

Our conformal symmetry breaking scale $\lambda $ is determined by
experimental data for the trajectory slope $W=4\lambda ^{2}$. This entails
the prediction $M_{\rho }=\sqrt{W/2}$ for the rho meson mass, which holds at
the few-percent level, and implies $\lambda =M_{\rho }/\sqrt{2}=0.55\ 
\mathrm{GeV}.$ In the baryon sector we similarly predict $M_{\Delta }=\sqrt{%
3W/2}=1.27\ \mathrm{GeV}$ for the isobar mass. The alternative determination
of $\lambda =M_{\Delta }/\sqrt{6}=0.50\ \mathrm{GeV}$ differs by less than
10\% from that in the rho channel and confirms the approximate universality
of $\lambda $ and $W=4\lambda ^{2}$. The slope $W$ is also related to the
QCD scale, estimated as $\Lambda _{\mathrm{QCD}}\simeq \sqrt{W/8}=\lambda
/\sqrt{2}\simeq 0.35~$GeV and close to the empirical value $\Lambda _{\mathrm{%
QCD}}$ $\simeq 0.33$~GeV (at hadronic scales with three active flavors)\cite%
{pdg}. The resulting trajectories match the experimental data well\cite%
{for07}, with two expected exceptions: the pion and nucleon come out too
heavy since the present, simplest version of our holographic dual lacks
chiral symmetry.

We conclude that our construction of a rather efficient and predictive QCD
gravity dual which generates the linear trajectories (\ref{bspec}),
including for the first time those empirically found in the baryon sector,
provides new support for the AdS/QCD program. It establishes the basis for
calculating gauge theory correlation functions and observables from the dual
mode solutions\cite{for07} on the basis of the AdS/CFT dictionary\cite{mal97}
and predicts new relations between the $\rho $ and $\Delta $ masses and
their trajectory slopes. Our method of reconstructing the dual mode dynamics
from spectral properties on the gauge theory side may be useful for other
applications as well.

\section*{Acknowledgements}

This work was supported by DAAD/CAPES, CNPq and FAPESP.

\end{document}